\documentclass[11pt]{article}
\usepackage{a4p}
\usepackage{amssymb}
\usepackage{epsfig}
\usepackage{cite}
\usepackage{pennames}
\usepackage{rotating}
\newcommand{\intLdt}  {183.1}

\newcommand{\dLstat}  {0.2}
\newcommand{\dLsys}   {0.4}

\newcommand{\rroots}  {188.635}  
\def\mrm{\mathrm}
\newcommand{\epem}{\ensuremath{\mathrm{e}^+\mathrm{e}^-}}
\newcommand{\ppbar}{\ensuremath{\mathrm{p}{\overline\mathrm{p}}}}
\newcommand{\mpmm}{\ensuremath{\mu^+\mu^-}}

\newcommand{\lplm}{\ensuremath{\ell^+\ell^-}}
\newcommand{\Zz}{\ensuremath{{\mathrm{Z}^0}}}
\newcommand{\WW}{\ensuremath{\mathrm{W}^+\mathrm{W}^-}}
\newcommand{\WpWm}{\ensuremath{\mathrm{W}^+\mathrm{W}^-}}
\newcommand{\Wp}{\ensuremath{\mathrm{W}^+}}
\newcommand{\Wpm}{\ensuremath{\mathrm{W}^\pm}}
\newcommand{\Wm}{\ensuremath{\mathrm{W}^-}}

\newcommand{\qq}{\ensuremath{\mathrm{q\overline{q}}}}
\newcommand{\QQ}{\ensuremath{\mathrm{qq}}}

\newcommand{\lnu}{\ensuremath{\ell\nu_\ell}}

\newcommand{\enu}{\ensuremath{\mathrm{e{\nu}_{e}}}}
\newcommand{\mnu}{\ensuremath{\mu{\nu}_{\mu}}}
\newcommand{\tnu}{\ensuremath{\tau{\nu}_{\tau}}}

\newcommand{\qqqq}{\ensuremath{\QQ\QQ}}
\newcommand{\qqln}{\ensuremath{\QQ\lnu}}

\newcommand{\WWqqln}{\ensuremath{\WW\rightarrow\QQ\lnu}}
\newcommand{\WWqqqq}{\ensuremath{\WW\rightarrow\QQ\QQ}}

\newcommand{\WWqqen}{\ensuremath{\WW\rightarrow\QQ\enu}}
\newcommand{\WWqqmn}{\ensuremath{\WW\rightarrow\QQ\mnu}}
\newcommand{\WWqqtn}{\ensuremath{\WW\rightarrow\QQ\tnu}}

\newcommand{\WWlnln}{\ensuremath{\WW\rightarrow\lnu\lnu}}

\newcommand{\Wenu}{\ensuremath{\epem \rightarrow \mathrm{W}\enu}}

\newcommand{\ZZqqll}{\ensuremath{(\Zz/\gamma)^*(\Zz/\gamma)^*\rightarrow\qq\lplm}}

\newcommand{\Zqq}{\ensuremath{\Zz/\gamma\rightarrow\qq}}

\newcommand{\Mz}{\ensuremath{M_{\mrm{Z}}}}
\newcommand{\Mw}{\ensuremath{M_{\mrm{W}}}}
\newcommand{\Mwqqqq}{\ensuremath{M_{\mrm{W}}^{\QQ\QQ}}}
\newcommand{\Mwqqln}{\ensuremath{M_{\mrm{W}}^{\QQ\lnu}}}
\newcommand{\Gw}{\ensuremath{\Gamma_{\mathrm{W}}}}
\newcommand{\DMw}{\ensuremath{\Delta M_{\mrm{W}}}}

\newcommand{\Opal}{\mbox{OPAL}}
\newcommand{\OPAL}{\mbox{OPAL}}

\newcommand{\Lund}{\mbox{L{\sc und}}}
\newcommand{\Jetset}{\mbox{J{\sc etset}}}
\newcommand{\JETSET}{\mbox{J{\sc etset}}}

\newcommand{\KoralW}{\mbox{K{\sc oralw}}}

\newcommand{\Excalibur}{\mbox{E{\sc xcalibur}}}

\newcommand{\grcff}{\mbox{grc4f}}
\newcommand{\Pythia}{\mbox{P{\sc ythia}}}

\newcommand{\Ariadne}{\mbox{A{\sc riadne}}}

\newcommand{\Herwig}{\mbox{H{\sc erwig}}}
\newcommand{\HERWIG}{\mbox{H{\sc erwig}}}

\newcommand{\GeV}{\ensuremath{\mathrm{GeV}}}
\newcommand{\MeV}{\ensuremath{\mathrm{MeV}}}
\newcommand{\Pfit}{\ensuremath{P_\mathrm{FIT}}}

\newcommand{\GeVcc}{\ensuremath{\mathrm{GeV}}}
\newcommand{\roots}{\ensuremath{\sqrt{s}}}
\newcommand{\rootsprime}{\ensuremath{\sqrt{s^\prime}}}

\newcommand{\Zgamma}{\ensuremath{\Zz/\gamma}}

\newcommand {\ra}         {\ensuremath{\rightarrow}}

\newcommand{\CC}{\mbox{{\sc CC03}}}
\newcommand{\mrec}{\ensuremath{m_{\mathrm{rec}}}}

\newcommand{\srec}{\ensuremath{\sigma_{\mathrm{rec}}}}
\def\etal{\mbox{{\it et al.}}}
\def\ie{\mbox{{\it i.e.}}}
\def\eg{\mbox{{\it e.g.}}}

\def\gappeq{\ensuremath{\mathrel{ \rlap{\raise.5ex\hbox{>}}
                      {\lower.5ex\hbox{\sim}}}}}
\def\lappeq{\ensuremath{\mathrel{ \rlap{\raise.5ex\hbox{<}}
                      {\lower.5ex\hbox{\sim}}}}}

\newcommand{\JPG}[3]  {J.\ Phys.\ \textbf{G#1} (#2) #3}
\newcommand{\PLB}[3]  {Phys.\ Lett.\ \textbf{B#1} (#2) #3}
\newcommand{\ZPC}[3]  {Z.\ Phys.\ \textbf{C#1} (#2) #3}
\newcommand{\EPC}[3]  {Eur.\ Phys.\ J.\ \textbf{C#1} (#2) #3}
\newcommand{\NIMA}[3] {Nucl.\ Instr.\ Meth.\ \textbf{A#1} (#2) #3}

\newcommand{\EP}[1]  {CERN-EP/{#1}}

\newcommand{\PRL}[3]  {Phys.\ Rev.\ Lett.\ \textbf{#1} (#2) #3}
\newcommand{\PRD}[3]  {Phys.\ Rev.\ \textbf{D#1} (#2) #3}
\newcommand{\NPB}[3]  {Nucl.\ Phys.\ \textbf{B#1} (#2) #3}

\newcommand{\CPC}[3]  {Comp.\ Phys.\ Comm.\ \textbf{#1} (#2) #3}
\newcommand{\Zzero}{\mbox{${\mathrm{Z}^0}$}}

\newcommand{\WWqqenu}{\mbox{\WW$\rightarrow$ \qq\enu}}
\newcommand{\WWqqmnu}{\mbox{\WW$\rightarrow$ \qq\mnu}}
\newcommand{\WWqqtnu}{\mbox{\WW$\rightarrow$ \qq\tnu}}

\newcommand{\Leptwo}{\mbox{L{\sc EP2}}}
\newcommand{\Lepone}{\mbox{L{\sc EP1}}}

\def\opalabbiendi{OPAL Collaboration, G.\ Abbiendi \etal}
\def\opalackerstaff{OPAL Collaboration, K.\ Ackerstaff \etal}
\def\opalalexander{OPAL Collaboration, G.\ Alexander \etal}
\def\opalakers{OPAL Collaboration, R.\ Akers \etal}
\def\opalacton{OPAL Collaboration, P.D.\ Acton \etal}

\begin{document}
\begin{titlepage}

\begin{center}{\Large   EUROPEAN ORGANISATION FOR NUCLEAR RESEARCH (CERN)
}\end{center}\bigskip
\begin{flushright}
       CERN-EP/2000-099   \\ 20th July 2000
\end{flushright}
\bigskip\bigskip\bigskip\bigskip\bigskip
\begin{center}
  {\huge\bf \boldmath Measurement of the  Mass and Width \\
    \vspace*{5mm}    
    of the W Boson in \epem\ Collisions at 189~GeV}
\end{center}
\bigskip\bigskip
\begin{center}
  {\LARGE The OPAL Collaboration}
\end{center}\bigskip\bigskip 
%
%
\begin{center}
  {\large \bf Abstract }
\end{center} 
The mass and width of the W boson are determined in $\epem$
collisions at LEP using 183~pb$^{-1}$ of data recorded at a 
centre-of-mass energy $\roots=189$~GeV with the \Opal\ detector.
The invariant mass distributions from 970 \WWqqqq\ and 1118 \WWqqln\
candidate events are used to measure 
the mass of the W boson, 
\begin{eqnarray*} 
  \Mw & = &80.451\pm0.076\mathrm{(stat.)} \pm0.049\mathrm{(syst.)}~\GeVcc.
\end{eqnarray*}
A direct measurement of the width of the W boson gives 
{\ensuremath{\Gw=2.09\pm0.18\mrm{(stat.)}\pm0.09\mrm{(syst.)}~\GeVcc}}.
The results are combined with previous \Opal\ results 
from 78~pb$^{-1}$ of data recorded with \roots\ from  161 to 183 GeV, 
to obtain: 
\begin{eqnarray*}
  \Mw & = & 80.432\pm0.066\mrm{(stat.)}\pm0.045\mrm{(syst.)}~\GeVcc, \\
  \Gw & = & 2.04\pm0.16\mrm{(stat.)}\pm0.09\mrm{(syst.)}~\GeVcc.
\end{eqnarray*}
The consistency of the direct measurement of $\Mw$ with that inferred
from other measurements of electroweak parameters provides an important
test of the Standard Model of electroweak interactions.    

\vspace*{1cm}

\begin{center}{\large
(Submitted to Physics Letters B)
}\end{center}

\end{titlepage}
\begin{center}{\Large        The OPAL Collaboration
}\end{center}\bigskip
\begin{center}{
G.\thinspace Abbiendi$^{  2}$,
K.\thinspace Ackerstaff$^{  8}$,
C.\thinspace Ainsley$^{  5}$,
P.F.\thinspace {\AA}kesson$^{  3}$,
G.\thinspace Alexander$^{ 22}$,
J.\thinspace Allison$^{ 16}$,
K.J.\thinspace Anderson$^{  9}$,
S.\thinspace Arcelli$^{ 17}$,
S.\thinspace Asai$^{ 23}$,
S.F.\thinspace Ashby$^{  1}$
% ...
}\end{center}\bigskip
$^{  1}$School of Physics and Astronomy, University of Birmingham,
Birmingham B15 2TT, UK
\newline
$^{  2}$Dipartimento di Fisica dell' Universit\`a di Bologna and INFN,
I-40126 Bologna, Italy
\newline
$^{  3}$Physikalisches Institut, Universit\"at Bonn,
D-53115 Bonn, Germany
\newline
% ...
$^{  5}$Cavendish Laboratory, Cambridge CB3 0HE, UK
\newline
% ...
$^{  8}$CERN, European Organisation for Nuclear Research,
CH-1211 Geneva 23, Switzerland
\newline
$^{  9}$Enrico Fermi Institute and Department of Physics,
University of Chicago, Chicago IL 60637, USA
\newline
% ...
$^{ 16}$Department of Physics, Schuster Laboratory, The University,
Manchester M13 9PL, UK
\newline
$^{ 17}$Department of Physics, University of Maryland,
College Park, MD 20742, USA
\newline
% ...
$^{ 22}$Department of Physics and Astronomy, Tel Aviv University,
Tel Aviv 69978, Israel
\newline
$^{ 23}$International Centre for Elementary Particle Physics and
Department of Physics, University of Tokyo, Tokyo 113-0033, and
Kobe University, Kobe 657-8501, Japan
\newline
% ...


\section{Introduction}
\label{intro} 

The LEP \epem\ collider at CERN provides an ideal environment for the
study of the properties of the gauge bosons of the Standard
Model (SM) of electroweak interactions\cite{bib:sm}. 
In the first
stage of its operation, LEP produced \epem\ collisions at 
centre-of-mass energies, $\roots$,  
within a few GeV of the $\Zzero$ resonance, allowing precise
measurements of the properties of the $\Zzero$ boson including its 
mass and fermionic couplings\cite{bib:lineshape,bib:lep-ewksum}. 
In the context of the SM
these measurements place constraints on the mass of the 
Higgs boson and provide an indirect determination of the mass of the W 
boson, \Mw.
Since 1996, the LEP collider has operated 
above the threshold for \Wp \Wm\ production (\Leptwo),
allowing measurements of the trilinear gauge 
boson couplings\cite{bib:tgc183} and a
direct measurement of \Mw.
When combined with the direct measurements of the top quark mass at the 
Tevatron\cite{bib:tevamtop}, measurements of \Mw\ enable further
constraints to be set on the mass of the Higgs boson via
electroweak radiative corrections \cite{bib:sirlin, bib:degressi}. 
Comparison between the direct measurements of the mass of the W boson
and the value determined indirectly from data recorded at
$\roots\approx\Mz$ provides an 
important test of the self-consistency of the Standard Model.
The direct measurement of $\Gw$ further tests the consistency
of the Standard Model.

The combination of direct measurements of \Mw\ from \Leptwo\ at 
\roots\ $\sim161-183$~GeV
\cite{bib:O-mw161,bib:O-mw172,bib:O-mw183,bib:A-mw183, 
bib:D-mw183,bib:L-mw183} 
and from hadron colliders \cite{bib:HAD-mw} currently 
give $\Mw = 80.419\pm0.056$~GeV\cite{bib:pdg00}. This direct measurement
is consistent with the indirect value obtained from lower energy data,
primarily measurements at $\roots\sim\Mz$, which give
$\Mw^{indirect} = 80.382\pm0.026$~GeV\cite{bib:lep-ewksum}.

The published OPAL 
measurements of $\Mw$ are based on approximately 78~pb$^{-1}$
of data. This paper describes a measurement of the mass of the W boson
using a further 183~pb$^{-1}$ of data recorded by \Opal\ during 1998 at
\roots$\sim189$~GeV.  This result is combined with previous \Opal\ 
measurements to give a direct measurement of the mass of the 
W boson with a total uncertainty of 79~MeV. 
It is expected that the ultimate LEP precision
on $\Mw$ will be approximately $30$~MeV 
when all data are included and the results of the four LEP experiments are
combined\cite{bib:LEP2YR}.

\section{Data and Monte Carlo Samples}

A detailed description of the \Opal\ detector can be found in 
\cite{bib:opaldet}.
The data sample used for this analysis corresponds to an 
accepted integrated luminosity, evaluated using small angle Bhabha
scattering events observed in the  
forward calorimeters\cite{bib:O-lumi}, of
$\intLdt\pm\dLstat \mathrm{(stat.)} \pm\dLsys
\mathrm{(syst.)}$~pb$^{-1}$. 
The luminosity weighted mean centre-of-mass energy for the data sample is
$\roots= \rroots \pm 0.040$~\GeV\cite{bib:energy189}.

\subsection{Monte Carlo samples}
 
A number of Monte Carlo generators are used to simulate the 
physics processes relevant to the studies presented in this paper. All
samples include a full simulation of the \Opal\
detector \cite{bib:GOPAL}. The main physics processes at \Leptwo\ can
be broken down into three main categories: four-fermion (4$f$) production,
including $\epem\rightarrow\WpWm\rightarrow 4f$ and 
$\epem\rightarrow\Zz\Zz\rightarrow 4f$, but excluding contributions from
multi-peripheral diagrams;
two-fermion production; and multi-peripheral two-photon mediated 
processes. 
For the measurement of the 
W boson mass and width, only the four-fermion processes 
and the two-fermion background process $\epem\rightarrow\Zqq$ 
play an important role.

The \KoralW\ program\cite{bib:KoralW142}, which uses matrix
elements calculated with \grcff\cite{bib:GRC4F}, is used to simulate 
the production of most four-fermion final states, $\epem\rightarrow 4f$. 
The main Monte Carlo samples are generated at \roots\ = 188.634~GeV 
with $\Mw=80.33$~\GeV. The samples were generated using 
the running width scheme for the Breit-Wigner distribution.
The four-fermion samples are divided into
final states which have contributions from processes involving 
the W-boson propagator and those which do not. 
In the invariant mass region close to $\Mw$, 
the four-fermion cross-section is dominated by  
doubly resonant W-pair production diagrams (CC03)\footnote{In this paper, the
doubly-resonant W pair production diagrams, {\em i.e.} $t$-channel
$\nu_{\mathrm{e}}$ exchange and $s$-channel \Zgamma\ exchange, are
referred to as ``\CC'', following the notation of \cite{bib:LEP2YR}.}.
Additional \KoralW\ four-fermion Monte Carlo samples are produced with
different centre-of-mass energies and with different values of \Mw.  
The four-fermion background from the process $\epem\rightarrow\epem\qq$
is simulated using the \grcff\ \cite{bib:GRC4F} generator.
The most important two-fermion background process, $\epem\rightarrow\Zqq$,
is simulated using \Pythia\cite{bib:Pythia}, with 
\HERWIG\ \cite{bib:HERWIG} used to assess possible systematic 
uncertainties. 

\section{\boldmath \WpWm\ Event Selection}

The event selections are described 
in~\cite{bib:O-xs189} and references therein.
The selections are sensitive to the leptonic \WWlnln,
semi-leptonic \WWqqln, and hadronic \WWqqqq\ final states.
Due to the presence of two unobserved  prompt neutrinos in the
$\WWlnln$ final state there is little sensitivity to $\Mw$
and the leptonic final state is therefore not used here.  

Semi-leptonic \WWqqln\ decays comprise $44\%$ of the total \WW\ 
cross-section. The event selection employs three multivariate 
relative likelihood discriminants, one
for each of the \WWqqen, \WWqqmn, and \WWqqtn\ final states.  The \WWqqen\
and \WWqqmn\ channels are characterised by two well-separated hadronic
jets, a high-momentum lepton and missing momentum due to the prompt
neutrino from the leptonic W decay.  The signature for the \WWqqtn\
channel is similar, with the exception that the $\tau$ lepton is
identified as an isolated, low-multiplicity jet typically consisting of
one or three tracks.  \WWqqln\ events are selected
with an efficiency of $87\%$ and a purity of $91\%$.  The dominant
backgrounds are \Zqq\ and four-fermion processes such as \Wenu\ and
$\epem\rightarrow\ZZqqll$. 

Hadronic \WWqqqq\ decays comprise 
$46\%$ of the total \WW\ cross-section and
are identified by requiring 
four energetic hadronic jets and little or no missing
energy. A preselection removes approximately $98\%$ of the dominant
background process, \Zqq.  A multivariate
relative likelihood discriminant is then employed to select the \WWqqqq\
candidates with an efficiency of $87\%$ and a purity of $77\%$.

After the selections are applied, 1546 \WWqqqq\ and 1246 \WWqqln\
candidate events remain, consistent with Standard Model expectations. Not all
events are used in the measurements described here. 
As discussed in Section~\ref{sec:mrec}, additional
cuts are applied to remove poorly reconstructed events 
and further reduce backgrounds.


\section{Measurement of the Mass and Width of the W Boson}
\label{Wmass}

The measurement of the mass and width of the W boson 
proceeds in two stages. Firstly the invariant masses of the W decay 
products are reconstructed on an event-by-event basis.  
Kinematic fits are applied to each 
selected \WWqqln\ and \WWqqqq\ event 
to improve the mass resolution. The reconstructed invariant 
mass spectra are then used to determine $\Mw$ and $\Gw$. 

Fits to the invariant mass spectra to obtain $\Mw$ are  
performed using three different techniques. The central results of 
this paper are obtained using a Monte Carlo 
reweighting technique~\cite{bib:LEP2YR} to fit the observed mass spectra
to obtain \Mw\ and \Gw. The W mass is also determined 
using two alternative methods, which are used as 
cross-checks.
In the first, an analytic fit to the measured mass 
spectrum uses an unbinned likelihood fit method to determine \Mw.  
To describe the signal shape, the fit uses a parametrisation based on a 
Breit-Wigner function \cite{bib:O-mw172}.  The second method uses a
convolution technique\cite{bib:D-mw183,bib:LEP2YR}.  
The three different fitting techniques have similar 
expected statistical sensitivities and similar estimated
systematic uncertainties. 

\subsection{Invariant mass reconstruction}
\label{sec:mrec}

The three methods for extracting \Mw\ use nearly identical
procedures to reconstruct the invariant mass of the W candidates. 
The description
here applies to the reweighting method.  Small variations 
relevant for the alternative analyses are discussed in 
Section~\ref{sec:alt-fits}.

In previous \Opal\ measurements of \Mw\ the tracks and clusters 
in selected \WWqqqq\ events were grouped into four jets 
using the Durham algorithm\cite{bib:durham}. For the results presented 
here, \WWqqqq\ events are grouped into either 
four or five jet topologies depending on the value
of $\ln(y_{45})$, where $y_{45}$ is the value of the Durham jet resolution
parameter at which the transition from 5 to 4 jets occurs.
Events with  $\ln(y_{45})>-5.6$ are treated
as five jets.  This separation allows for the possibility of
hard gluon radiation from one of the quarks and is found
to improve the expected statistical sensitivity to $\Mw$ 
by approximately $5\%$ compared with treating all events as four jets. 
The improvement in statistical sensitivity comes from events where,
under the four jet hypothesis, a gluon jet from one W boson is 
combined with a quark jet from a different W boson.
In the semi-leptonic decay channels, \WWqqln, the lepton candidate 
is removed and the hadronic part of the event is reconstructed as 
two jets. The division of $\WWqqln$ events into two and three jet
would not improve the mass resolution. 
After the association of tracks and
clusters into jets, corrections derived from the Monte Carlo
simulation  are applied to the measured jet momenta to account for
double counting from particles which deposit energy in more than
one sub-detector\cite{bib:MT}.

The invariant masses of the two W bosons can be determined 
directly from the reconstructed momenta of the observed decay products. 
However, the mass resolution is limited by the relatively large uncertainty 
on the measured energies of the jets, $\sigma_{E}/E \approx 20\%$, 
rather than by uncertainties on the measured jet directions.
For this reason the use of a kinematic fit which imposes the four 
constraints of energy and momentum conservation\cite{bib:kfit} 
(4-C fit) significantly 
improves the invariant mass resolution. The improvements mainly arise
from the energy in event being constrained to the well measured
centre-of-mass energy.
The output of the 4-C fit consists of two reconstructed
masses per event, one for each W boson in the final state.
The resolution of the kinematic fit can be improved by imposing a 
further constraint that the masses of the two reconstructed W boson 
candidates are equal (5-C fit), yielding a single reconstructed
mass per event. For $\WWqqln$ events the three unmeasured variables
corresponding to the neutrino momentum means that the effective
number of constraints in the semi-leptonic is two, giving a 2-C fit. 
For the results presented in this paper, 
the mass reconstruction is performed using the 
5-C fit for $\WWqqqq$ events and the 2-C fit for $\WWqqln$ events.

A common kinematic fitting algorithm is used for $\WWqqen$, $\WWqqmn$ 
and $\WWqqqq$ events.  The fitted mass is obtained using an 
iterative $\chi^2$-minimisation procedure where
the constraints in the kinematic fit are implemented 
using Lagrange multipliers. The presence of initial state radiation (ISR) 
is neglected. The corrected jet momenta, their associated 
errors, and the measured jet masses are input to the kinematic fit.  
The errors on the measured jet momenta are parameterised
by expressions derived from Monte Carlo studies, which are functions of
the visible energy and polar angle of the jet.  
The jet masses are fixed to their measured values. This is found to
improve the mass resolution from the fit compared to treating jets 
as massless and also to reduce the bias in the fitted mass distribution.   

The mass reconstruction for $\WWqqtn$ events is different.
In the \WWqqtn\ system, the absence of a measurement of the 
tau lepton energy means that the W mass information is determined 
entirely by the hadronic system.  The absence of a measurement
of the tau lepton energy reduces the effective number of constraints
in the kinematic to one. Previously\cite{bib:O-mw183} the reconstructed mass 
in \WWqqtn\ events was obtained from the invariant mass of the 
jet-jet system, scaled by the ratio of the beam energy to the sum of the 
jet energies. For the results presented here an analytic formula which 
reproduces the results of the 1-C kinematic fit is used. 

For \WWqqen\ and  \WWqqmn\ events the lepton direction is
taken to coincide with the direction of the track associated with the
electron or muon candidate.  The energy is estimated from the associated
electromagnetic calorimeter cluster for electrons and from the momentum of
the track for muons. Unassociated electromagnetic clusters close to the
lepton track, consistent with being from final state radiation (FSR), are
included in the energy calculation. For each event a single mass 
is determined from a 2-C kinematic fit. In addition to the fitted mass, 
\mrec, the error on the fitted mass, \srec, and the
$\chi^2$ fit probability, \Pfit,  are calculated. Events with 
$\mrec>65$~GeV  and $\Pfit>0.001$ are retained. 
About half of the $\WWqqtn$ events selected as either 
$\WWqqen$ or $\WWqqmn$ fail these. The mass information is recovered by 
treating events with $\mrec<65$~GeV  or $\Pfit<0.001$ as \WWqqtn\ events. 
In addition, about $4\%$ of selected $\WWqqen$ and $\WWqqmn$ events 
have the identified lepton beyond the effective
tracking acceptance of the OPAL 
detector, within $20^\circ$ of the beam axis, 
in which case the lepton energy is either poorly
measured or not measured at all. These events are included in the 
$\WWqqtn$ channel. The numbers of events used in each of 
the \WWqqln\ channels are given in Table~\ref{tab:evno}.

\begin{figure}
  \epsfxsize=\textwidth
  \epsffile{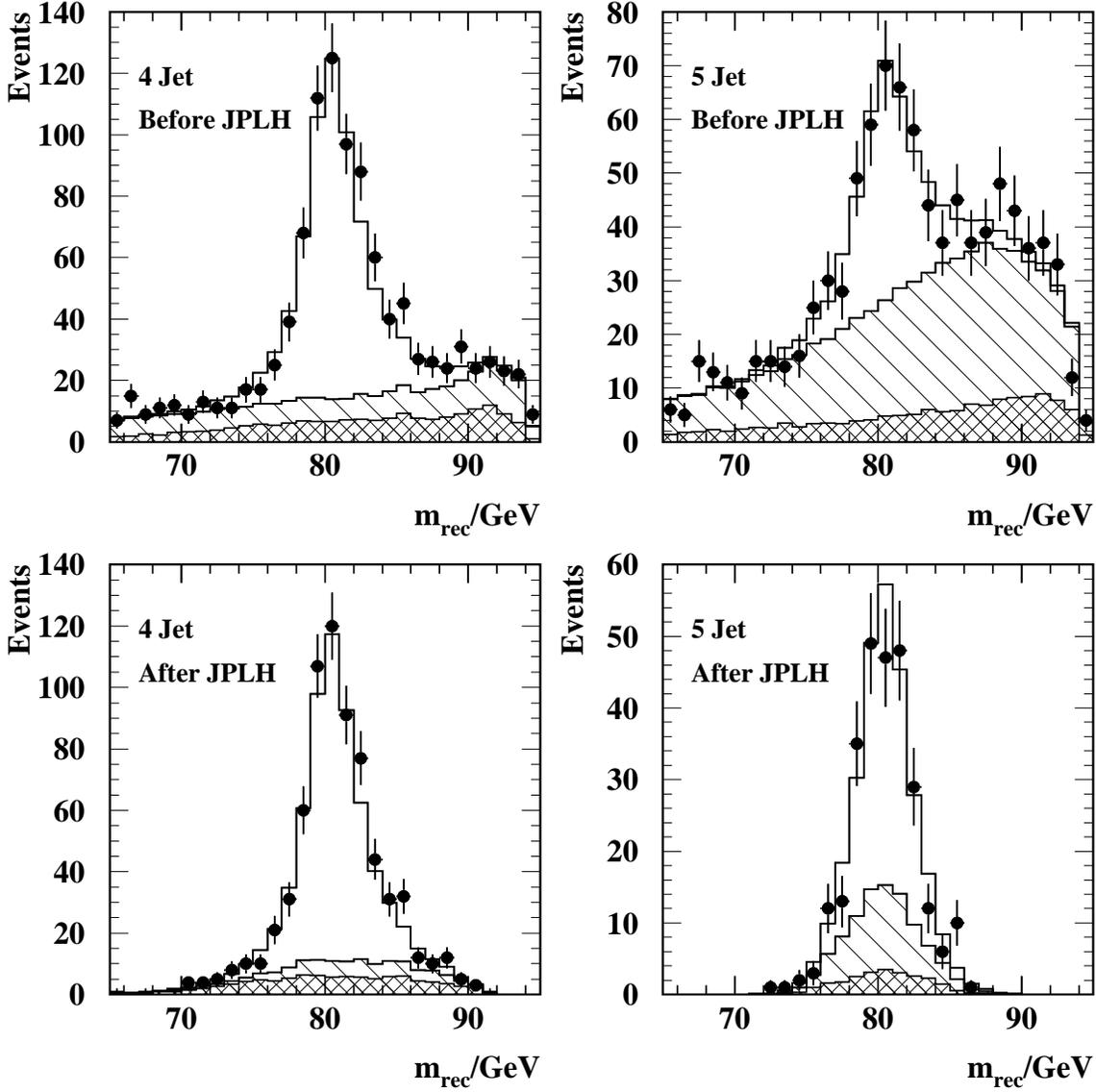}
  \caption{ The reconstructed mass, $\mrec$, distributions for \WWqqqq\ 
    events divided into four and five jet samples.  The top two plots
    show the distributions before the jet-pairing likelihood (JPLH) cut.
    The lower two plots show the mass distributions which are used
    to determine \Mw, {\em i.e.} after the cut on JPLH. 
    The points correspond to the
    \Opal\ data and the histograms to the Monte Carlo predictions. The 
    contribution from the non-WW background is shown as the cross-hatched 
    histogram and the combinatorial background is indicated by
    the singly-hatched histogram.}
\label{fig:jet-pairing}
\end{figure}

The situation for \WWqqqq\ events is complicated by the fact that
there are three possible assignments of four jets to 
the two W bosons. For events reconstructed as five jets there
are ten possible assignments of the jets to the W bosons.
Incorrect combinations contain little or no information
on the mass of the W boson. The incorrect jet-pairings result in a 
combinatorial background. For each \WWqqqq\ event, 
three (or ten) kinematic fits are performed, corresponding to the possible 
jet-pairings.  To eliminate poorly reconstructed events and reduce
backgrounds, only combinations which give a successful 5-C kinematic
fit with a resulting $\chi^2$
fit probability $\Pfit>0.01$ and $\mrec\: >\:65$~GeV are considered.
A multivariate relative likelihood discriminant similar to  that described in
\cite{bib:O-mw183} is employed to pick out a single combination for each
event and reduce combinatorial background.  Different variables
are used for the jet-pairing likelihoods  for the four and five
jet cases. For four jet events, the combination used in the fit
is selected on the basis of two variables: the difference between the
two fitted masses from the 4-C fit and the sum of the di-jet opening angles.
In the five jet sample four variables are used: the 5-C fit mass,
the difference between the two fitted masses from the 4-C fit, the 
minimum opening angle between the jets in the system assigned as 
$\Wpm\rightarrow\qq{g}$ and the cosine of the polar angle of the 
reconstructed $\Wpm\rightarrow\qq{g}$ system. For events reconstructed 
as four (five) 
jets, the combination corresponding to the largest jet-pairing likelihood
is
retained provided it has a likelihood output exceeding $0.40 (0.42)$.  
Monte Carlo studies indicate that in $89\%$ ($70\%$) of the
surviving signal events, the selected combination corresponds to the
correct jet-pairing. The number of surviving events
in the \WWqqqq\ channel is given in Table~\ref{tab:evno}.
Figure \ref{fig:jet-pairing} shows the reconstructed mass distributions in
both four and five jet channels before and after the jet-pairing
likelihood cuts.

For events without ISR the average $\mrec$ resolution
(as defined in \cite{bib:O-mw183}) 
for the correct jet pairing in $\WWqqqq$ events is 1.7~GeV. For 
$\WWqqln$ events the average $\mrec$ resolution is 2.4~GeV, 2.8~GeV and
3.4~GeV in the $\WWqqen$, $\WWqqmn$ and $\WWqqtn$ channels respectively.  

\subsection{Extraction of the W mass and width}
\label{ssec:RWfit}

The Monte Carlo reweighting technique is used to provide the
central results of this paper.  The W boson mass and width are 
measured by directly comparing the reconstructed mass spectra in the 
data to Monte Carlo mass spectra corresponding to different 
values of \Mw\ and \Gw. A likelihood fit is then used to extract 
\Mw\ and \Gw\ by determining which Monte Carlo spectrum best describes 
the data. The Monte Carlo spectra for arbitrary (\Mw, \Gw) 
are obtained using the reweighting technique described 
in~\cite{bib:O-mw172}. In previous \Opal\ 
publications, $\WW$ Monte Carlo samples generated using only CC03 diagrams  
were reweighted using the ratio of
Breit-Wigner functions. For this analysis, \KoralW\
$\epem\rightarrow 4f$ reference samples generated with 
$(\Mw=80.33~\GeV$, $\Gw=2.093~\GeV)$
are reweighted to $(\Mw^\prime,\Gw^\prime)$ using the ratio of
Breit-Wigner functions. The reference samples used for the 
reweighting include only final states which have contributions from
diagrams involving the W propagator. This procedure is found to give a good 
approximation to the more exact treatment of using the full four-fermion 
matrix elements, introducing a bias of less than $5$~MeV in the $\WWqqen$
channel and less than 2~MeV in the other channels. Using the ratio of
Breit-Wigner functions rather than four-fermion matrix elements
results in a much faster fit. 

\begin{figure}
  \epsfxsize=\textwidth
  \epsffile{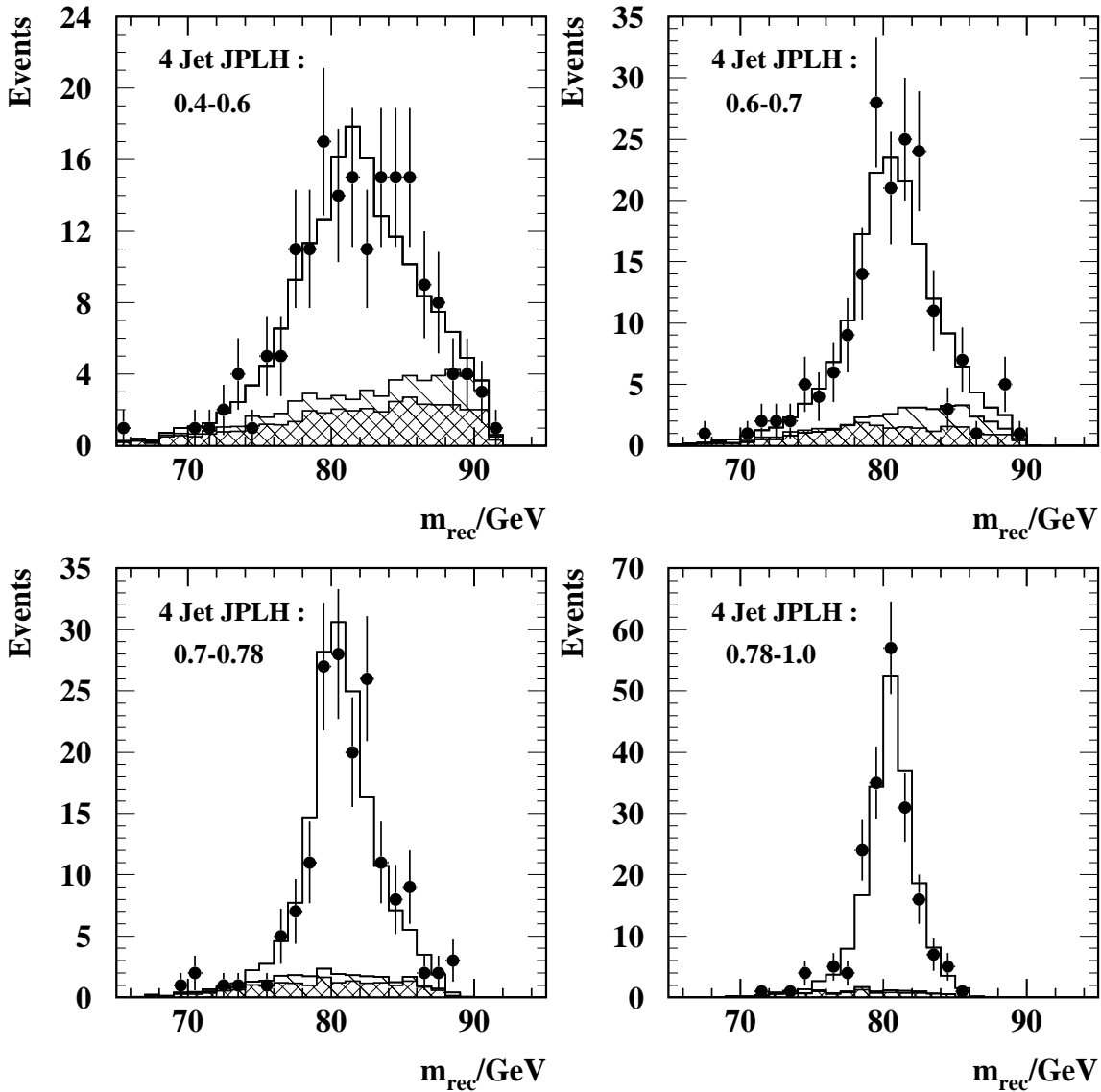}
  \caption{ The reconstructed mass distributions for the four jet \WWqqqq\ 
    sample divided into four bins of the output of the 
    jet-pairing likelihood (JPLH) discriminant. 
    The points correspond to the
    \Opal\ data and the histograms to the Monte Carlo predictions. The 
    contribution from the non-WW background is shown as the cross-hatched 
    histogram and the combinatorial background is indicated by
    the singly-hatched histogram.}
\label{fig:mass-res}
\end{figure}

The mass spectra for background events are taken from Monte Carlo and are 
assumed to be independent of \Mw\ and \Gw.  The main 
sources of background are $\epem\rightarrow\Zqq$ 
and four-fermion processes.
The background mass distributions are normalised to the 
expected number of background events.  
The reweighted signal spectra are then normalised such that the 
total number of signal plus background events corresponds to the 
observed number of events. 
This is done separately for the \WWqqqq, \WWqqen, \WWqqmn\ and 
\WWqqtn\ channels, with the \WWqqqq\ channel split into four and five jet 
topologies. In addition, the \WWqqln\ channels are divided into four 
subsamples according to the error on the reconstructed invariant mass, \srec. 
This division 
gives a larger weight to events with reconstructed masses which are known with 
better precision (\ie\ small \srec ) and reduces the expected 
statistical uncertainty on the fitted W mass from the \WWqqln\ channels
by approximately $5\%$. 
In the \WWqqqq\ channels, the width of the reconstructed mass distribution 
is dominated by the intrinsic width of the W. Consequently a 
similar subdivision into bins of $\srec$ does not improve the 
\Mw\ sensitivity.  However, jet-pairings which give a large 
jet-pairing likelihood  are more likely to be 
correct and have a better mass resolution
as shown in Figure \ref{fig:mass-res}.
For this reason, the $\WWqqqq$ events reconstructed as four jets 
are subdivided into four bins of jet-pairing likelihood, resulting in 
a $7\%$ improvement in the statistical sensitivity to $\Mw$. No subdivision
is performed for the five jet events.

A binned log-likelihood fit to the \mrec\ distributions of the data is 
performed in the range $\mrec > 65~\GeV$.  The log-likelihood function is 
identical to that used  previously\cite{bib:O-mw183}.
A log-likelihood curve is determined 
separately for each channel.  For the \WWqqln\ channels, the results are 
obtained by adding the log-likelihood curves separately determined from each 
channel in each bin of $\srec$. 
For the \WWqqqq\ channel, the results are 
obtained by adding the likelihood curves obtained from the five 
jet events to the likelihood curves obtained from the four 
bins of jet-pairing likelihood for events reconstructed as four jets.

\begin{figure}
  \epsfxsize=\textwidth
  \epsffile{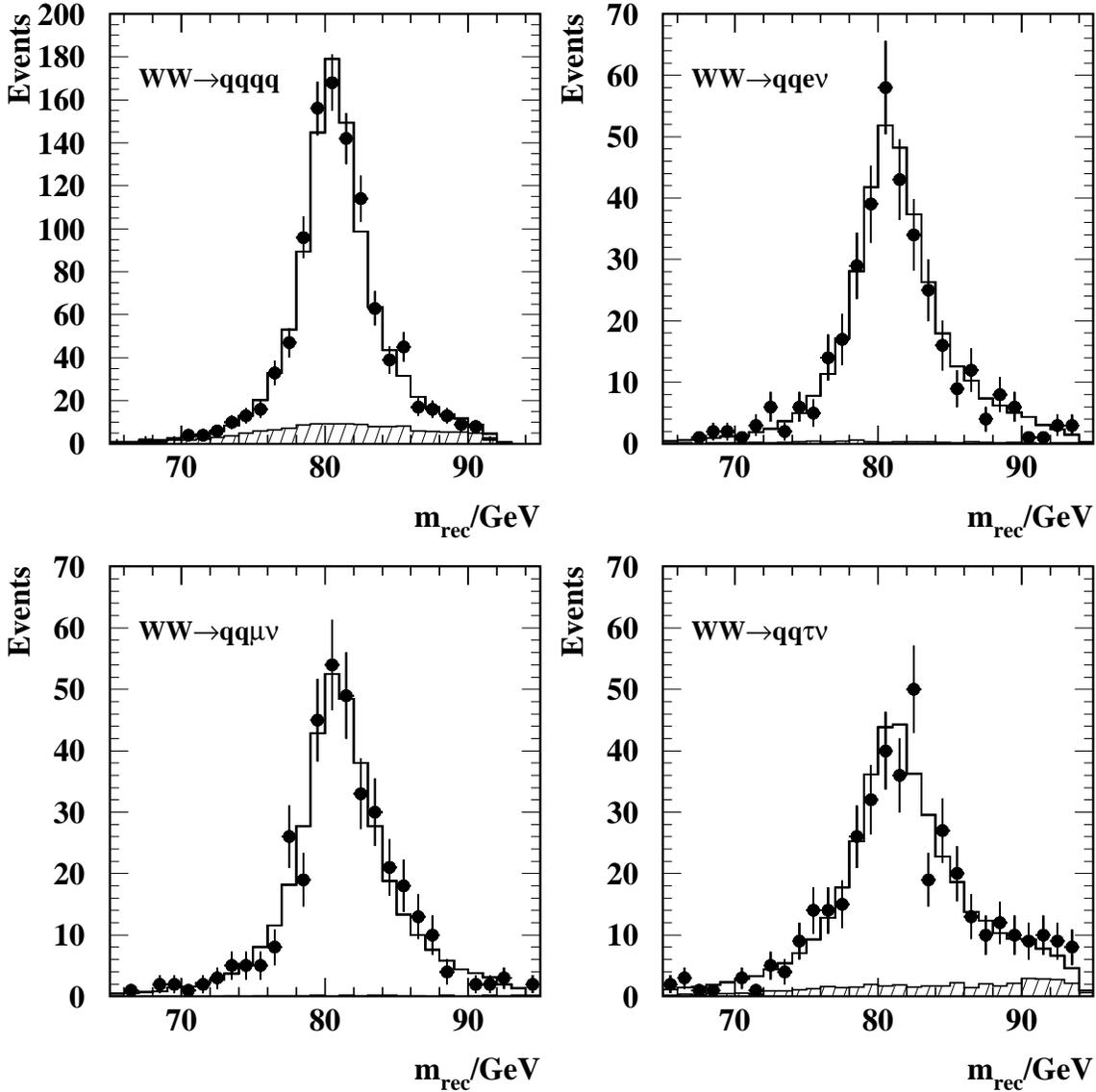}
  \caption{ The reconstructed invariant mass distribution for the \WWqqqq, 
    \WWqqen, \WWqqmn\ and \WWqqtn\ samples.  The points correspond to the
    \Opal\ data and the histograms to the reweighted Monte Carlo spectra
    corresponding to the fitted masses. 
    The non-WW background contribution is indicated by the hatched histogram. }
\label{fig:rwdata}
\end{figure}

Two types of fit are performed.  In the one parameter fit, \Gw\ is constrained
by the SM \mbox{relation\cite{bib:LEP2YR}},
\begin{eqnarray} 
 \Gw & = & 3 G_{\mathrm{F}}\Mw^3
(1+2\alpha_{\mathrm{S}}/3\pi)/(2\sqrt{2}\pi)  ,  
      \label{eqn:smmwgw}
\end{eqnarray}
and only \Mw\ is determined. 
The results of this fit for each channel are given in 
Table~\ref{tab:mw-rew} and displayed in Figure~\ref{fig:rwdata}.  The combined
result is discussed in Section~\ref{sec:results}.  In the two parameter fit, 
both \Mw\ and \Gw\ are determined simultaneously.  

In the reweighting method the fitted parameters are expected to be
unbiased since any offsets in the reconstructed mass
introduced in the analysis are implicitly 
accounted for in the Monte Carlo reconstructed mass spectra 
used in the reweighting procedure.
This is verified using several Monte Carlo samples generated at various
\Mw\ and \Gw.  In addition, tests using a large ensemble of Monte Carlo 
subsamples, each corresponding to 183~pb$^{-1}$ and including background 
contributions, are used to verify for each channel separately and for all 
channels combined, that the measured fit errors accurately reflect the 
root-mean-squared  
spread of the residual distribution for both the \Mw\ and \Gw\ fits.  
Since $\Mw$ in the Monte Carlo corresponds to the running width definition,
so does the fitted mass. 
The expected statistical error on the W mass from the combination 
of the $\qqln$ and $\qqqq$ channels is $76\pm1$~MeV, where the 
weights given to the two channels are determined by the both statistical
and systematic error contributions. The quoted uncertainty on the 
expected statistical error is from Monte Carlo statistics.

\subsection{Alternative fit methods}
\label{sec:alt-fits}

\subsubsection{Breit-Wigner fit}

The Breit-Wigner method is analogous to that described  in~\cite{bib:O-mw183}.
It employs an unbinned maximum-likelihood fit to the reconstructed mass 
spectrum using an analytic Breit-Wigner function to describe the 
mass spectrum from $\epem\rightarrow\WW$. Due to
initial-state radiation,  the reconstructed mass spectrum is 
asymmetric.  For the $\WWqqln$ channels a 
relativistic Breit-Wigner function, with different widths above and below the 
peak, gives a satisfactory description of the \mrec\ lineshape. 
The fitting function used is
\begin{eqnarray}
  S( m_{\mrm{rec}} ) & = & A\frac{m^{2}_{\mrm{rec}} \, \Gamma_{+(-)}^{2}}
   {(m^{2}_{\mrm{rec}}-m_{0}^{2})^{2} + m^{2}_{\mrm{rec}} \, 
   \Gamma_{+(-)}^{2}},
\end{eqnarray}
where $\Gamma_{+(-)}$ is the width assumed for all \mrec\ above (below) the
peak centred at $m_{0}$.  This empirical choice of fitting function provides
an adequate description for samples up to ten times the integrated luminosity
of the data.  The widths, $\Gamma_{+(-)}$, are fixed to values 
determined from fits to \WW\ signal Monte Carlo samples.
Different widths are obtained for each $\WWqqln$ decay channel.  
The shapes of the background distributions and the background fractions are 
determined from Monte Carlo. The background fractions are held constant in the 
fit. The fit is restricted to the range $70 < \mrec < 88$~GeV.

In the $\WWqqqq$ channel events are divided into four and five jet samples
which are fitted separately. The division is made on the basis of the
5-C kinematic fit probability. 
Events are reconstructed as five jets if any of the ten possible jet-pairings 
in the five jet assignment gives a fit probability which is greater than
twice that of highest probability of the three jet-pairings for the
four jet hypothesis. For both four jet and five jet samples, 
the fitting function $S(m_{\mrm{rec}})$ is multiplied by $G(m_{\mrm{rec}}) 
= \exp[-(m_{0}-m_{\mrm{rec}})^2/2\sigma^2]$. This empirical choice 
provides a good description of the reconstructed mass spectra in the fit range 
for $\WWqqqq$ events. The value of $\sigma$ is determined from Monte Carlo.
For the four jet \WWqqqq\ sample either one or two jet-pairings are
used following the procedure described in 
\cite{bib:O-mw172}. In the five jet sample, 
a jet-pairing likelihood is used. Here the variables used in the
likelihood are those used for the reweighting fit 
described in Section \ref{sec:mrec}, with the exception that the 
5-C fit mass is not included.

In contrast to the procedure employed for the reweighting method, 
the \WWqqln\ events are not divided into subsamples according to \srec\
nor are the \WWqqqq\ events divided into subsamples according to the
jet-pairing likelihood. However, \WWqqtn\ events are sub-divided into
fully-leptonic and semi-leptonic decays of the tau lepton. 

The fitted mass,  $m_{0}$, must be corrected for offsets not accounted for in
the fit, \eg\ from initial-state radiation and event selection. The correction 
is determined using fully simulated Monte Carlo samples generated at 
different values of \Mw\ with the expected background contributions 
included and is found to depend linearly on \Mw. 
The results from the \WWqqqq\ and \WWqqln\ channels, after correction, are
given in Table~\ref{tab:alt-fits}. The expected statistical error
on the combined W mass measurement is $78\pm2$~MeV.

\subsubsection{Convolution fit}

The convolution method\cite{bib:LEP2YR,bib:D-mw183} attempts to exploit all
available information by constructing a likelihood curve for each selected
event.  The likelihood is calculated using a functional form
\begin{displaymath}
  \mathcal{L}(\Mw, \mrec ) = p_s \mathcal{P}_{s}( \Mw, \mrec ), 
\end{displaymath}
where $p_s$ is the probability of a candidate event being a true signal
event and $\mathcal{P}_s$ is the probability function for 
$\Mw$ given the observed reconstructed mass, $\mrec$.  
The function $\mathcal{P}_{s}( \Mw , \mrec )$ is defined as
\begin{displaymath}
  \mathcal{P}_s( \Mw , \mrec )=\mrm{BW}(\Mw, m, s^\prime) 
                       \otimes \mrm{ISR}(s, s^\prime)
                       \otimes \mrm{R}(m, \mrec),
\end{displaymath}
where $\mrm{BW}(\Mw, m, s^\prime)$ is the relativistic Breit-Wigner function 
for producing off-shell W bosons with mass $m$ for a W mass of $\Mw$
including the effects of phase space. 
$\Gw$ is fixed relative to $\Mw$ using Equation~\ref{eqn:smmwgw}.
The radiator function, $\mrm{ISR}(s,s^\prime)$, is used to account for 
the effects of ISR (on a statistical basis). 
The inclusion of this term accounts for the fact that 
in the kinematic fit the observed 
energy is constrained to $\roots$, where in reality the effective 
centre-of-mass energy, $\rootsprime$, depends on the amount of ISR. 
Finally, $\mrm{R}(m, \mrec)$ is the
probability density function  relating the experimentally reconstructed 
mass, $\mrec$, to the average of the two true masses, $m$, of the 
off-shell W bosons in the event.

In the \WWqqln\ channels, the full error information on the fitted mass of
each event is used in an unbinned maximum likelihood fit. 
 In a previous \Opal\ publication
\cite{bib:O-mw183}, the event probability function, $\mrm{R}(m, \mrec)$, 
was assumed to be
Gaussian with its central value and standard deviation taken from a 2-C
kinematic fit.  In this analysis, the 2-C 
fit is modified, replacing the equal W mass constraint by one in which the
masses are individually constrained to a given W boson mass, $m$.  The
$\chi^2$ distribution from this fit is converted into the event
probability density function, $\mrm{R}(m, \mrec)$, which is non-Gaussian.

In the \WWqqqq\ channel, all events are forced into a five jet topology.  A
relative likelihood, constructed from the difference between the
reconstructed masses in a 4-C kinematic fit, is used to select on average
about three jet-pairings per event thereby  
reducing significantly
the combinatorial background.  In the Monte Carlo 
$92\%$ of selected events the correct 
jet-pairing combination is used in the fit.
The convolution is performed in two dimensions, namely
the reconstructed masses
of the two W bosons, with joint
probability density functions corresponding to each jet-pairing obtained
from the 4-C kinematic fit. The probability density functions for the
different combinations are added to form an event probability density
function\cite{bib:D-mw183}.

In both channels the log-likelihood curves from each selected event are
summed to yield a single curve from which a fitted mass is determined.  
The fitted mass is corrected for offsets not accounted for in
the fit in the manner described for the Breit-Wigner fit.  
The results from the \WWqqqq\ and \WWqqln\ channels,
after all corrections, are given in Table~\ref{tab:alt-fits}.  The expected
statistical error on the combined W mass measurement is 
$72\pm2$~MeV. The total combined expected error, statistical and systematic, 
from the convolution method is slightly smaller than that obtained from
the reweighting fit. However, at present the reweighting fit is the only 
method in which a simultaneous fit for $(\Mw,\Gw)$ is implemented.
For this reason, and for consistency with previous OPAL publications,
the reweighting method is retained for the results given in this publication. 

\section{Systematic Uncertainties}
\label{sec:syst}

The systematic uncertainties for the measurement of $\Mw$ using
the reweighting method 
are estimated as described below and summarised in Table~\ref{tab:sys}. 
The contributions from each of the sources 
are added in quadrature to yield the total systematic uncertainty. 
All contributions were evaluated for both the \Mw\ and \Gw\
determinations. For the alternative analyses, the 
systematics are estimated similarly and yield comparable results.
The determination of the individual systematic uncertainties is
described below.

\subsection{Beam energy}

  The average LEP beam energy is currently known with a precision of 
  $20$~MeV \cite{bib:energy189} which leads to a systematic uncertainty
  of 16~MeV on $\Mw$ and 2~MeV on $\Gw$. 
  The RMS spread in the LEP centre-of-mass 
  energy of $237\pm12$~MeV\cite{bib:energy189} and the average
  difference between the electron and positron beam energies results in 
  negligible impact on the measurement of $\Mw$. The Monte Carlo
  does not include the intrinsic beam energy spread. This introduces a
  bias of $+10$~MeV to the value of $\Gw$ obtained from the 
  reweighting fit.  A $-10$~MeV correction is applied and the
  size of the correction taken as an additional systematic 
  uncertainty on the measurement $\Gw$.

    
\subsection{Initial state radiation}

  The main \KoralW\ Monte Carlo samples include an 
  ${\cal{O}}(\alpha^3)$ treatment of initial
  state radiation (ISR). The systematic error associated with ISR is 
  estimated by reweighting the generated \KoralW\
  events to correspond to an ${\cal{O}}(\alpha^2)$ or 
  ${\cal{O}}(\alpha)$ treatment 
  of ISR using the matrix elements calculated in \KoralW. 
  The resulting shifts from the comparison of ${\cal{O}}(\alpha^3)$ to 
  ${\cal{O}}(\alpha)$ are less than 2~MeV and are neglected.  

\subsection{\boldmath Full ${\cal{O}}(\alpha)$ corrections}

In the process $\epem\rightarrow 4f\gamma$ ISR diagrams dominate. 
The accuracy of the treatment of ISR in the \KoralW\ generator is 
sufficient for this analysis. However the \KoralW\ generator does not include
all ${\cal{O}}(\alpha)$ effects. For example, interference between
ISR and FSR graphs and photon radiation from the W bosons are not implemented.
Recent calculations using the double-pole approximation indicate 
that possible mass biases of order 10~MeV could arise\cite{bib:racoon}.
However, these estimates were based on generator level studies which
compared mass distributions from the full ${\cal{O}}(\alpha)$ 
treatment with the corresponding Born level treatment which does
not include any final state radiation.  
To obtain a proper estimate of the potential mass bias when
applied to the experimental measurement it would be
necessary to use the full OPAL detector simulation and mass reconstruction
procedure.  Currently this is not possible since the necessary
Monte Carlo programs are not yet available.
For this paper possible systematic biases from the full 
${\cal{O}}(\alpha)$ treatment of photon radiation are not included
since the size of the effect is significantly smaller than the
current statistical precision of the $\Mw$ measurement. 

\subsection{\boldmath Hadronisation in $\Wpm\rightarrow\qq$ decays}

  The \KoralW\ Monte Carlo samples taken as references in the extraction of
  $\Mw$ use the \Lund\ string model as implemented in 
  \Jetset\ for the simulation of the hadronisation\footnote{In this
  paper hadronisation refers to the process whereby the quarks produced
  in W decays produce the hadrons observed in the detector
  including the development of the parton shower.} of 
  $\Wpm\rightarrow\qq$ decays. The \Jetset\ model was tuned to 
  \OPAL\ hadronic data recorded at the $\Zz$ resonance\cite{bib:jsettune}.
  The model was tuned to describe event shape variables and inclusive
  particle production rates. To assess a possible systematic error
  from the uncertainties in the tuning of the \Jetset\ model,
  the \Jetset\ parameters $\sigma_{q}$, $b$, 
  $\Lambda_{\mrm{QCD}}$ and $Q_{0}$ are each varied in turn 
  by $\pm1\sigma$ about their tuned values at the $\Zz$\cite{bib:jsettune}. 
  The maximum resulting biases to the fitted values of $\Mw$ from these
  high statistics samples are $30$~MeV 
  and 20~MeV in the $\qqln$ and $\qqqq$ channels respectively. 
  In addition, an earlier \OPAL\ tune of the \Jetset\ model 
  is used. This version was tuned to describe global event shape variables
  alone\cite{bib:oldjsettune}. 
  The resulting differences in the fitted value of \Mw\ using the earlier
  \OPAL\ tune are
  $(-33\pm14)$~MeV  and $(-30\pm12)$~MeV 
  in the $\WWqqqq$ and $\WWqqln$ channels respectively.

  As a further test of
  hadronisation uncertainties,
  the \Jetset\ string model is replaced by the \Herwig\ cluster model. For 
  this comparison, two samples were generated using the 
  same \WW\ final states and differing only in the 
  hadronisation modelling.
  The resulting fitted mass values from the reweighting method 
  from \JETSET\ and \HERWIG\ are compared.
  Differences (\JETSET-\HERWIG) of $(-18\pm14)$~MeV and 
  $(-11\pm12)$~MeV are obtained for the  $\WWqqln$  and $\WWqqqq$ channels.
  A similar comparison is made using the \Ariadne\ colour dipole 
  model\cite{bib:ariadne}. Differences (\JETSET-\Ariadne) 
  of $(+15\pm 14)$~MeV and $(+6\pm 12)$~MeV are obtained for the  
  $\WWqqln$  and $\WWqqqq$ channels.

  On the basis of these comparisons
  systematic uncertainties of 30~MeV are assigned to the
  W mass measurements from both $\WWqqln$ and $\WWqqqq$. The
  hadronisation uncertainty is taken to be fully correlated between the
  two channels. For the measurement of $\Gw$ the corresponding
  systematic errors in the $\WWqqln$ and $\WWqqqq$ channels
  are 25~MeV and 60~MeV respectively.

\subsection{Detector calibration and simulation}

  The effects of detector mis-calibrations and deficiencies in the 
  Monte Carlo are investigated by varying the jet and lepton energy 
  scales over reasonable ranges. 
  The ranges used for the systematic variations are dependent 
  on polar angle and are determined from detailed comparisons of
  data and Monte Carlo utilising both data recorded at 189~GeV and 
  $3.1\:\mathrm{pb}^{-1}$ of data collected at $\roots\approx\Mz$
  during 1998.  
  From $\epem\ra\epem$ events the
  electro-magnetic calorimeter energy scale is known to  
  $0.3\%$. Similarly from $\epem\ra\Zzero\ra\mpmm$ events, the uncertainty
  on momentum scale from the tracking detectors is determined to be $0.3\%$. 
  These scale uncertainties dominate the detector related systematics 
  in the $\WWqqen$ and $\WWqqmn$ channels, and are
  14~MeV and 16~MeV respectively. 
  Uncertainties on energy (momentum) resolution and angular resolution are 
  also evaluated but have much smaller impact.
  Uncertainties on the jet energy scale are determined from 
  $\epem\ra\Zzero\ra\qq$ events to be $0.2\%$--$1.0\%$ depending on the 
  polar angle. 
  The dominant effect of the jet energy scale uncertainty
  is a 6~\MeV\ uncertainty in the $\WWqqtn$ channel. In the $\WWqqqq$
  channel the sensitivity to the overall jet energy
  scale is greatly reduced due to the kinematic fit. Here the dominant
  effect is from the global uncertainty in the jet polar
  angle scale, {\em i.e.} the relative length-to-width scale of the 
  detector, resulting in a 3~MeV uncertainty on the W mass.
  The linearity of the jet energy scale is checked using 
  $\epem\ra\Zzero\ra\qq$ events with a clear three-jet topology. 
  For the measurement of $\Gw$ the main systematic error arises
  from  uncertainties  in the jet energy scale and resolution and is  47~MeV
  in both $\WWqqln$ and $\WWqqqq$ channels.

\subsection{Four-fermion modelling}
  Possible systematic effects associated with the modelling of the
  four-fermion final state, including interference between \WW\ diagrams
  and other four-fermion processes, are investigated by comparing the
  \grcff\ and the \Excalibur\cite{bib:EXCALIBUR} generators.  
  The comparison is performed using the ratio of the 
  \Excalibur\ to \grcff\ four-fermion
  matrix elements to reweight the \KoralW\ Monte Carlo sample. 
  This procedure avoids introducing an additional statistical error 
  associated with generating additional Monte Carlo.
  The resulting shift in the reconstructed mass is $1\pm1$~MeV in the
  $\WWqqln$ channel and  $0\pm1$~MeV in the $\WWqqqq$ channel. 
  Similar results are obtained when the \Excalibur\ Monte Carlo is reweighted
  using the ratio of the \grcff\ to \Excalibur\ four-fermion
  matrix elements. Consequently no additional systematic uncertainty 
  is assigned. 
  The corresponding systematic uncertainties on the measurement of 
  $\Gw$, of order 30~MeV, are determined by the statistical precision of
  the comparisons.  
  In the $\WWqqen$ channel an additional 5~MeV uncertainty is assigned,
  associated with using the Breit-Wigner functions in the reweighting fit. 

\subsection{Background treatment}
   Uncertainties in the modelling of background processes 
   mainly arise from the simulation of the 
   $\Zqq$ background in the 
   $\WWqqtn$ and $\WWqqqq$ channels. 
   A number of systematic checks of the normalisation
   and shape of the $\Zqq$ background are performed.
   The background normalisation in the reweighting fit 
   is varied by one standard deviation,
   using the uncertainties evaluated in~\cite{bib:O-xs189}. 
   The \Herwig\ cluster model is used as an alternative 
   to the \Lund\ string model used in \Pythia.
   In addition, for the \WWqqqq\ channel, data taken at 
   $\roots\approx\Mz$, scaled by $( 189\:\GeV / \Mz )$, are also substituted 
   for the \Zqq\ background.  
   The sum in quadrature of the above systematic shifts leads to the 
   assignment of a background systematic error of 
   6~MeV in the $\WWqqqq$ channel and 8~MeV in the
   $\WWqqln$ channel. The systematic error on $\Gw$ due to uncertainties
   in background modelling is 25~MeV and 41~MeV in the 
   $\WWqqln$ and $\WWqqqq$ channels respectively arising from differences 
   between \Jetset\ and \Herwig\ and uncertainties on the accepted
   background cross section.

\subsection{Monte Carlo statistics}
   The finite statistics of the Monte Carlo samples used in the 
   reweighting procedure result in an additional uncertainty of $10$~MeV 
   on the W mass determined separately in the \WWqqqq\ and \WWqqln\ channels.
   The corresponding systematic uncertainties for the W width measurement
   are 51~MeV.

\subsection{Final state interactions (FSI)}
  A significant bias to the apparent W mass measured in the  
  \WWqqqq\ channel could arise if the hadronisation of the two W bosons
  is not independent\cite{bib:LEP2YR,bib:wwqcd}. 
  At \Leptwo\ energies, the decay length for W bosons is about $0.1$~fm, 
  significantly less than the typical hadronisation scale, 
  approximately 1~fm.  The hadronisation of the two W bosons in 
  the \WWqqqq\ channel may have space-time overlap allowing possible
  final state interactions between the decay products of 
  different W bosons.
  Two possible sources of final state interactions, both leading to
  non-independent hadronisation, have been widely considered: 
  Colour Reconnection (CR) and Bose-Einstein Correlations (BEC). 
  Whether these effects play a significant role 
  at \Leptwo\ has yet to be experimentally determined. As a result,
  to assign a systematic error related to possible final state interactions
  it is necessary to rely on phenomenological models. 
  The strategy adopted here 
  is to consider only those models for CR and BEC which are 
  consistent with both \Lepone\ and \Leptwo\ data. 

  The CR systematic error refers to possible biases in the 
  W mass measurement arising from QCD reconnection effects 
  in the non-perturbative phase.
  Reconnection effects in the perturbative phase, {\em i.e.} multiple 
  gluon exchange between quarks from different W bosons, 
  have been shown to be small\cite{bib:sjk}. 
  To investigate the systematic biases 
  originating from non-perturbative CR
  effects, the models implemented in the \Pythia\ and \Ariadne\ 
  \cite{bib:ariadne} Monte Carlo generators are studied. 
  In both cases the results from the mass fit to a Monte Carlo sample
  including CR effects are compared to the results obtained from
  the corresponding sample without colour reconnection.
  The results are summarised in Table~\ref{tab:cr}.   
  The largest biases are seen in the \Ariadne\ 2 and \Ariadne\ 3
  models\footnote{Following the notation used 
  in~\cite{bib:O-cr183,bib:qqg}.}. However,
  these models have been strongly disfavoured by studies 
  of three jet events in \Lepone\ \Zz\ data\cite{bib:qqg}. 
  For this reason the current \Ariadne\ 
  implementation of CR is not used to assign a systematic 
  uncertainty.  Of the remaining models considered, the
  SK I model\cite{bib:sjk} produces the 
  largest bias. 
  The SK Monte Carlo samples use the \Jetset\ tune of \cite{bib:jsettune},
  where the parton shower cut-off, which directly influences the 
  behaviour of the CR model\cite{bib:sjk}, was set to 1.9~GeV.
  The strength parameter $k_i$
  (also called $\rho$) of the SK I model was set to
  0.9, corresponding to $35\%$ of events
  being reconnected. A systematic uncertainty of $66$~MeV on 
  \Mw\ is assigned. The corresponding uncertainty on \Gw\ is  $68$~MeV.

  As is the case in $\Zz$ decays, BEC 
  between like-sign charged pions are observed in $\Wpm$ decays at 
  \Leptwo\cite{bib:O-bec183}. These correlations are not implemented in 
  the standard Monte Carlo programs used in this analysis. 
  Correlations between pions from the  same W do not 
  affect significantly the global jet properties and are 
  unlikely to bias the $\Mw$
  measurement.  If BEC exist 
  between pions from  different W bosons in $\WWqqqq$ events,
  significant biases to the reconstructed value of $\Mw$ may
  arise.  To investigate possible systematic biases from BEC, a sample of 
  \WWqqqq\ events is produced using the \Pythia\ Monte Carlo 
  generator which includes BEC  
  as described in \cite{bib:pyt-be}. 
  For the evaluation of the systematic error the BEC model
  $BE_{32}$ of \cite{bib:pyt-be} 
  with $\lambda=1.0$ and $R=0.42$~fm is used. 
  The difference in the fitted W mass 
  between a sample generated with 
  BEC between different W bosons and a sample generated with BEC only
  between particles from the same W is $+67\pm14$~MeV (different-same).  
  The resulting systematic uncertainties on $\Mw$ and $\Gw$
  from BEC in the $\WWqqqq$ channel 
  are $67$~MeV and $39$~MeV respectively.

\subsection{Fit procedure}
   A comparison of the results from the reweighting method with those
   from the Breit-Wigner and convolution fits, summarised in Table 
   \ref{tab:alt-fits}, 
   is used to test for residual biases in the reweighting method.
   The expected RMS of the differences of the results from the three
   methods are determined using an ensemble of Monte Carlo 
   subsamples including background, 
   each corresponding to an integrated luminosity of $183$~pb$^{-1}$.
   For each subsample the difference in the fitted \Mw\ 
   determined using the reweighting method and that 
   determined using each of the other methods is calculated for the \WWqqqq\ 
   and \WWqqln\ channels. These differences have mean values 
   consistent with zero and RMS values of approximately $54$~MeV and
   $41$~MeV ($62$~MeV and $84$~MeV) when comparing the reweighting
   fits to the Breit-Wigner (convolution) fits in the \WWqqln\ and \WWqqqq\
   samples, respectively.  Since the alternative analyses yield 
   results consistent with those obtained using the default reweighting 
   analysis, no additional systematic error is 
   assigned.

\section{Results}
\label{sec:results}

\begin{figure}[htb]
  \epsfxsize=\textwidth
  \epsfbox[0 0 567 680]{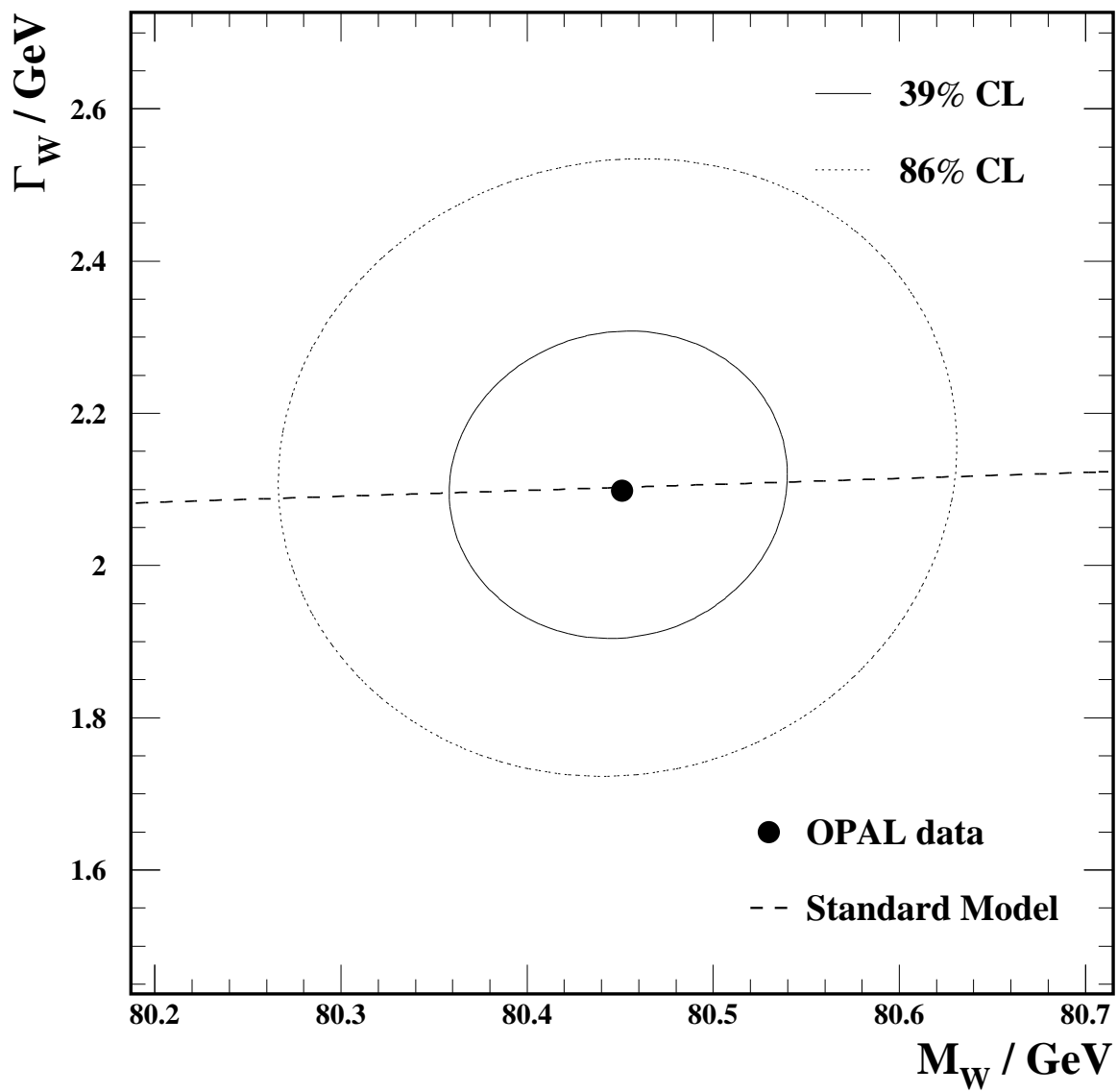}
  \caption
  {The $39\%$ and $86\%$ contour levels of the two parameter fit using the
    reweighting method including systematic contributions.
    The Standard Model relation between $\Mw$ and \Gw, given in Equation 
    \ref{eqn:smmwgw}, is shown by the dashed line. }
 \label{fig:errcontour}
\end{figure}

For the reweighting method described in Section~\ref{ssec:RWfit}, the results 
of a simultaneous fit to \Mw\ and \Gw\ from the combined \WWqqqq\ and \WWqqln\ 
event samples are
\begin{eqnarray*}
 \Mw & = & 80.451 \pm 0.076 (\mrm{stat.}) \pm 0.050 (\mrm{syst.})\, \GeV, \\
 \Gw & = &  2.09  \pm 0.18 (\mrm{stat.}) \pm 0.09  (\mrm{syst.})\, \GeV.
\end{eqnarray*}
The correlation coefficient between \Mw\ and \Gw\ is 0.04.  For this fit, the
central values are determined by adding the log-likelihood curves from the
\WWqqqq\ and \WWqqln\ channels including the effects of systematic 
uncertainties.  The likelihood contours for
this fit are displayed in Figure~\ref{fig:errcontour}.

A one parameter fit for the mass is performed by constraining the width using 
Equation \ref{eqn:smmwgw} to give 
$\Mw =80.402\pm0.104 (\mrm{stat.}) \pm 0.101 (\mrm{syst.})$~GeV in 
the \WWqqqq\ channel, and $\Mw = 80.478\pm 0.104 (\mrm{stat.}) \pm 0.038 
(\mrm{syst.})$~GeV in
the \WWqqln\ channel.  The combined result, taking into account 
the correlated systematics between the \WWqqqq\ and \WWqqln\ channels, is
\begin{eqnarray*}
   \Mw & = & 80.451 \pm 0.076 (\mrm{stat.}) \pm 0.049 (\mrm{syst.})\, \GeV.
\end{eqnarray*}
Due to the larger systematic errors in the \WWqqqq\ channel, mainly
from FSI, it carries a reduced weight of $0.36$ in the combination. 
The combined \WWqqqq\ and \WWqqln\ 
results from the alternative analyses, after all corrections, are for the 
Breit-Wigner fit, 
$\Mw=80.436\pm 0.078 (\mrm{stat.}) \pm 0.050 (\mrm{syst.})$~GeV, and for the 
convolution fit, 
$\Mw=80.378 \pm 0.073(\mrm{stat.})\pm 0.051(\mrm{syst.})$~GeV.

\subsection{Combination with previous data}

The results presented in this paper, based on 183~pb$^{-1}$ of data, 
are combined with previous \Opal\ measurements of \Mw\ from direct 
reconstruction at $\roots\approx172$~GeV\cite{bib:O-mw172} and
at $\roots\approx183$~GeV\cite{bib:O-mw183}.
The results in the $\WWqqln$ and $\WWqqqq$ channels are 
combined independently taking into account systematic uncertainties 
which are correlated between data from 
the different centre-of-mass energies, giving:
\begin{eqnarray*}
  \Mw (\WWqqln) & = & 80.441 \pm 0.086 (\mrm{stat.}) \pm 0.034 (\mrm{syst.}) \pm 0.017 (\mrm{lep})\, \GeV, \\
  \Mw (\WWqqqq) & = & 80.409 \pm 0.093 (\mrm{stat.}) \pm 0.034 (\mrm{syst.}) \pm 0.016 (\mrm{lep}) \pm 0.094 (\mrm{fsi})\, \GeV.
\end{eqnarray*} 
where the uncertainties from the LEP   
beam energy (lep) and from final state interactions (fsi) are
quoted separately. In the combination the systematic uncertainties for the 
colour reconnection and Bose-Einstein correlations of the previous
data were set to those presented in this paper.  
 
The difference between the fitted \Mw\ in the \WWqqqq\ and \WWqqln\ channels 
is $\DMw \equiv ( \Mwqqqq - \Mwqqln ) = -0.032 \pm 0.127 (\mrm{stat.}) 
\pm 0.047 (\mrm{syst.})~\GeV$.
A significant non-zero value for \DMw\ could indicate that 
FSI effects are biasing the value of \Mw\ determined from \WWqqqq\ events.
The systematic error on the quoted value of $\DMw$ 
does not include contributions from CR/BEC effects.  For the
$\DMw$ measurement the hadronization systematic uncertainties in the 
$\WWqqqq$ and $\WWqqln$ channels are taken to be uncorrelated;
a conservative choice in the absence of a good estimate of the true 
correlation coefficient.

The \Mw\ results from direct reconstruction in the $\WWqqln$ and
$\WWqqqq$ channels are combined with the \Opal\ measurement of \Mw\ from the 
\WW\ production cross-section at $\roots\approx161$~GeV\cite{bib:O-mw161}. 
The combination is made assuming that the mass measurements from direct 
reconstruction and from the threshold 
cross-section are uncorrelated, apart from the uncertainty associated with the
LEP beam energy, which is taken to be fully correlated.  The direct 
reconstruction measurements from the $\WWqqln$ and $\WWqqqq$ channels
from three different centre-of-mass energies are combined using the 
full covariance matrix. 
The combined result is
\begin{displaymath}
        \Mw = 80.432 \pm 0.066 (\mrm{stat.}) \pm 0.032 (\mrm{syst.}) 
                     \pm 0.028 (\mrm{fsi})   \pm 0.017 (\mrm{lep})\, \GeV.
\end{displaymath} 

The measurement of \Gw\ presented in this paper is
combined with the previous \Opal\ measurements 
from direct reconstruction\cite{bib:O-mw172,bib:O-mw183}
to give
\begin{displaymath}
        \Gw = 2.04 \pm 0.16 (\mrm{stat.}) \pm 0.09 (\mrm{syst.})\, \GeV.
\end{displaymath}   
For the earlier data, with relatively low statistics, 
the measured statistical error is correlated with the measured width.  
Consequently, to avoid biasing the combination, the separate width 
measurements are weighted using the expected statistical error
rather than the measured statistical error.

\section{Summary}

Using the $183$~pb$^{-1}$ of data recorded by the \Opal\ detector at a mean 
centre-of-mass energy of approximately 189~GeV, a total of 2088 \WWqqqq\ and 
\WWqqln\ candidate events are used in a fit constraining \Gw\ using
Equation \ref{eqn:smmwgw} to obtain a direct measurement of the W boson mass,
{\ensuremath{\Mw=80.451\pm0.076\mrm{(stat.)}\pm0.049\mrm{(syst.)}}}~GeV, while
a second fit is used to determine directly the width of the W boson,
{\ensuremath{\Gw=2.09\pm0.18 \mrm{(stat.)}\pm0.09\mrm{(syst.)}}}~GeV.

The results described in this paper are combined with the previous \Opal\ 
results from data recorded at $\roots \approx 
161-183$~GeV\cite{bib:O-mw161,bib:O-mw172,bib:O-mw183}. 
From this combined data sample the W boson mass is
determined to be
\begin{displaymath}
        \Mw = 80.432 \pm 0.066 (\mrm{stat.}) \pm 0.045 (\mrm{syst.}) \, \GeV.
\end{displaymath} 
This result is consistent with both the other direct measurements and
the indirect value inferred from fits to electroweak 
data\cite{bib:lep-ewksum}. It should be noted
that the systematic error is dominated by the uncertainties
arising from hadronisation and final state interactions.
For future LEP combinations
these uncertainties will need to be reduced to benefit from the 
ultimate LEP combined statistical uncertainty of about 25~MeV.    

The result for the W boson width is combined with the previous \Opal\ result 
from data recorded at $\roots \approx 172-183$~GeV to obtain
\begin{displaymath}
        \Gw = 2.04 \pm 0.16 (\mrm{stat.}) \pm 0.09 (\mrm{syst.}) \, \GeV.
\end{displaymath} 
This measurement of \Gw\ is in good agreement with the 
Standard Model expectation and with other 
direct measurements at LEP\cite{bib:D-mw183,bib:L-mw183,bib:A-mw189}
and at the Tevatron\cite{bib:CDFGW}.

%
%
%
%
\appendix
\par
\noindent
{\Large\bf Acknowledgements}

\par
\noindent
We particularly wish to thank the SL Division for the efficient operation
of the LEP accelerator at all energies
 and for their continuing close cooperation with
our experimental group.  We thank our colleagues from CEA, DAPNIA/SPP,
CE-Saclay for their efforts over the years on the time-of-flight and trigger
systems which we continue to use.  In addition to the support staff at our own
institutions we are pleased to acknowledge the  \\
Department of Energy, USA, \\
National Science Foundation, USA, \\
Particle Physics and Astronomy Research Council, UK, \\
Natural Sciences and Engineering Research Council, Canada, \\
Israel Science Foundation, administered by the Israel
Academy of Science and Humanities, \\
Minerva Gesellschaft, \\
Benoziyo Center for High Energy Physics,\\
Japanese Ministry of Education, Science and Culture (the
Monbusho) and a grant under the Monbusho International
Science Research Program,\\
Japanese Society for the Promotion of Science (JSPS),\\
German Israeli Bi-national Science Foundation (GIF), \\
Bundesministerium f\"ur Bildung und Forschung, Germany, \\
National Research Council of Canada, \\
Research Corporation, USA,\\
Hungarian Foundation for Scientific Research, OTKA T-029328, 
T023793 and OTKA F-023259.\\

%
%
\clearpage

\renewcommand{\arraystretch}{1.0}
\begin{table}[htbp]
  \begin{center}
    \begin{tabular}{|l|ccc|} \hline
      Channel & Observed & Expected & Purity \\ \hline
      \WWqqqq\ (4-jet) &  701 &  694 & $83\%$ \\
      \WWqqqq\ (5-jet) &  269 &  277 & $91\%$ \\
      \WWqqenu\        &  350 &  376 & $98\%$ \\
      \WWqqmnu\        &  365 &  373 & $99\%$ \\
      \WWqqtnu\        &  403 &  403 & $89\%$ \\ \hline
      Combined         & 2088 & 2123 & $91\%$ \\ \hline
    \end{tabular}
  \end{center}
  \caption[foo]{
    Numbers of events used in the W mass and width determination for each 
    channel and all channels combined.  Only events surviving the cuts 
    described in Section~\ref{sec:mrec} are included. The \WWqqtnu\ numbers
    include some events initially selected as either $\WWqqen$ or $\WWqqmn$.}
  \label{tab:evno}
\end{table} 
\vspace*{1.0cm}

\begin{table}[htbp]
  \begin{center}
    \begin{tabular}{|l|cc|} \hline
      Channel        & Measured \Mw/GeV  & Expected error/GeV \\ \hline
      \WWqqen        & $80.375 \pm 0.175$ & $0.164$ \\
      \WWqqmn        & $80.513 \pm 0.163$ & $0.168$ \\
      \WWqqtn        & $80.594 \pm 0.227$ & $0.220$ \\ 
      \WWqqqq(4-jet) & $80.424 \pm 0.114$ & $0.112$ \\ 
      \WWqqqq(5-jet) & $80.290 \pm 0.257$ & $0.230$ \\ \hline
      \WWqqln        & $80.478 \pm 0.104$ & $0.104$ \\ 
      \WWqqqq        & $80.402 \pm 0.104$ & $0.100$ \\ \hline
    \end{tabular}
  \caption{ Results using the reweighting method for the fit from 
    $183$~pb$^{-1}$ of data taken at $\roots\approx189$~GeV for each of the 
    channels separately and for the combined \WWqqln\ channel.
    Equation \ref{eqn:smmwgw} is used to constrain \Gw\ to $\Mw$.
    The expected errors are estimated using an ensemble of Monte Carlo 
    subsamples.
    The errors obtained in the data are consistent with the RMS spread
    of the errors obtained from the Monte Carlo subsamples.}
  \label{tab:mw-rew}
  \end{center}
\end{table}
\vspace*{1.0cm}

\begin{table}[htbp]
  \begin{center}
    \begin{tabular}{|l|c|c|c|} \hline
         & Reweighting & Breit-Wigner fit & Convolution fit \\
 Channel & Measured \Mw/GeV  &  Measured \Mw/GeV & Measured \Mw/GeV \\ \hline
 \WWqqln & $80.48\pm0.10\pm0.04$ & $80.47\pm0.11\pm0.04$ & $80.38\pm0.11\pm0.04$ \\
 \WWqqqq & $80.40\pm0.10\pm0.10$ & $80.38\pm0.11\pm0.10$ & $80.37\pm0.09\pm0.09$ \\ \hline
    \end{tabular}
  \caption{Comparison of $\Mw$ fit results obtained using the three different
    fitting techniques for the \WWqqln\ and \WWqqqq\ channels separately.
    In each case the first uncertainty is statistical and the 
    second systematic.  
    The observed differences are 
    compatible with the expected RMS differences between the results from
    the different fits.}
  \label{tab:alt-fits}
  \end{center}
\end{table}
\vspace*{1.0cm}

\begin{table}[htbp]
  \begin{center}
    \begin{tabular}{|l||c|c||c|c|} \hline
      Systematic errors &\multicolumn{2}{|c||}{ \Mw } &
                         \multicolumn{2}{|c|}{ \Gw }     \\
        ~~~~~~~~(MeV) 
                                 &\qqqq&\qqln&\qqqq&\qqln \\ \hline
      Beam Energy                &  16 &  16 &  10 &  10  \\
      Hadronisation              &  30 &  30 &  60 &  25  \\
      Jet energy/Resolution      &   4 &   5 &  47 &  47  \\
      Lepton energy/Resolution   & $-$ &   9 & $-$ &   9  \\
      Four Fermion               & $-$ & $-$ &  33 &  29  \\
      Background                 &   6 &   8 &  41 &  25  \\
      MC statistics              &  10 &  10 &  51 &  51  \\ \hline
      Sub-total                  &  36 &  38 & 106 &  84  \\ \hline
      Colour Reconnection        &  66 & $-$ &  68 & $-$  \\ 
      Bose-Einstein Correlations &  67 & $-$ &  39 & $-$  \\ \hline
      Total systematic error     & 101 &  38 & 132 &  84  \\ \hline
    \end{tabular}
  \end{center}
  \caption[foo]{
    Summary of the systematic uncertainties for the fit results.  For the fits
    to determine \Mw, \Gw\ is constrained to its Standard Model relation.
    The uncertainties are given separately for fits to the \WWqqqq\
    and \WWqqln\ samples. Uncertainties from ISR  
    give negligible contributions to both \Mw\ and \Gw. The uncertainty 
    related to the Monte Carlo models of the four-fermion process
    give is negligible for the measurement of $\Mw$.}
  \label{tab:sys}
\end{table} 
\vspace*{1.0cm}

\begin{table}[htbp]
  \begin{center}
   \begin{tabular}{|l|r|r|} \hline
    ~~~~CR~Model & Mass shift (MeV)  & Width shift (MeV)  \\ \hline
    ~~~~~~~~SK I           & $+66 \pm 8$~~~~~~ & $+68  \pm 22$~~~~~~ \\
    ~~~~~~~~SK II          & $ +3 \pm 8$~~~~~~ & $+20  \pm 22$~~~~~~ \\
    ~~~~~~~~SK II$^\prime$ & $+10 \pm 8$~~~~~~ & $+53  \pm 21$~~~~~~ \\
    ~~~~~~~~AR 2           & $+85 \pm 8$~~~~~~ & $+128 \pm 22$~~~~~~ \\
    ~~~~~~~~AR 3           & $+140\pm 10$~~~~~~ & $+309 \pm 21$~~~~~~ \\ \hline
    \end{tabular} 
 \caption{ The predicted shifts in the fitted mass and width from the 
  \WWqqqq\ channel  for various colour-reconnection (CR) models.  
  The numbers correspond to the fitted mass (width) using the reweighting 
  method in the reconnected sample minus the fitted mass (width) 
  for the corresponding sample without CR. 
  The errors are statistical only.}
  \label{tab:cr}
 \end{center}
\end{table}
%
%
\newpage

\end{document}